\documentclass{llncs}

\usepackage{graphicx}
\usepackage{epsf}

\def\Figure#1#2#3{
\begin{figure}[htb]
\epsfxsize=#2
\begin{center}
\includegraphics[width=#2]{#1}
\end{center}
\caption{#3}
\label{#1}
\end{figure}}

\usepackage{amssymb}
\usepackage{graphicx}

\usepackage{url}
\urldef{\mailsa}\path|alexander.gamkrelidze@tsu.ge|
\newcommand{\keywords}[1]{\par\addvspace\baselineskip
\noindent\keywordname\enspace\ignorespaces#1}

\begin{document}

\mainmatter

\title{Algorithms for Low-Dimensional Topology}

\author{Alexander Gamkrelidze}

\authorrunning{}

\institute{{Department of Computer Science,\\
Tbilisi State University,\\
Building XI, Room 355,\\
Tbilisi, Georgia}
\mailsa\\
}

\maketitle

\begin{abstract}
In this article, we re-introduce the so called "Arkaden-Faden-Lage" (AFL for short) representation of knots in 3 dimensional space introduced by Kurt Reidemeister and show how it can be used to develop efficient algorithms to compute some important topological knot structures. In particular, we introduce an efficient algorithm to calculate holonomic representation of knots introduced by V. Vassiliev and give the main ideas how to use the AFL representations of knots to compute the Kontsevich Integral.

The methods introduced here are to our knowledge novel and can open new perspectives in the development of fast algorithms in low dimensional topology.

\end{abstract}


\section{Introduction}

The main goal of this article is to show how an old and idea can be analyzed and applied in new light to solve actual mathematical problems. As an example we use the so-called "Arkaden-Faden-Lage" (AFL for short) representation of knots in order to develop efficient algorithms to solve two important actual problems: the holonomic description of knots and the computation of the Kontsevich Integral for knots. The idea to use the AFL representations in low dimensional topology is published in \cite{Gam1}.

The AFL representation of knots was first introduced by Kurt Reidemeister. It is a modification of the Gauss representation of knots and can have more effective applications in practice. 

In \cite{Gauss}, Gauss showed that each knot can be projected into a plane so that it can be decomposed in two non-self-crossing parts. As an example, fig. \ref{Trefoil-1}(a) shows such a representation of a trefoil. 

\Figure{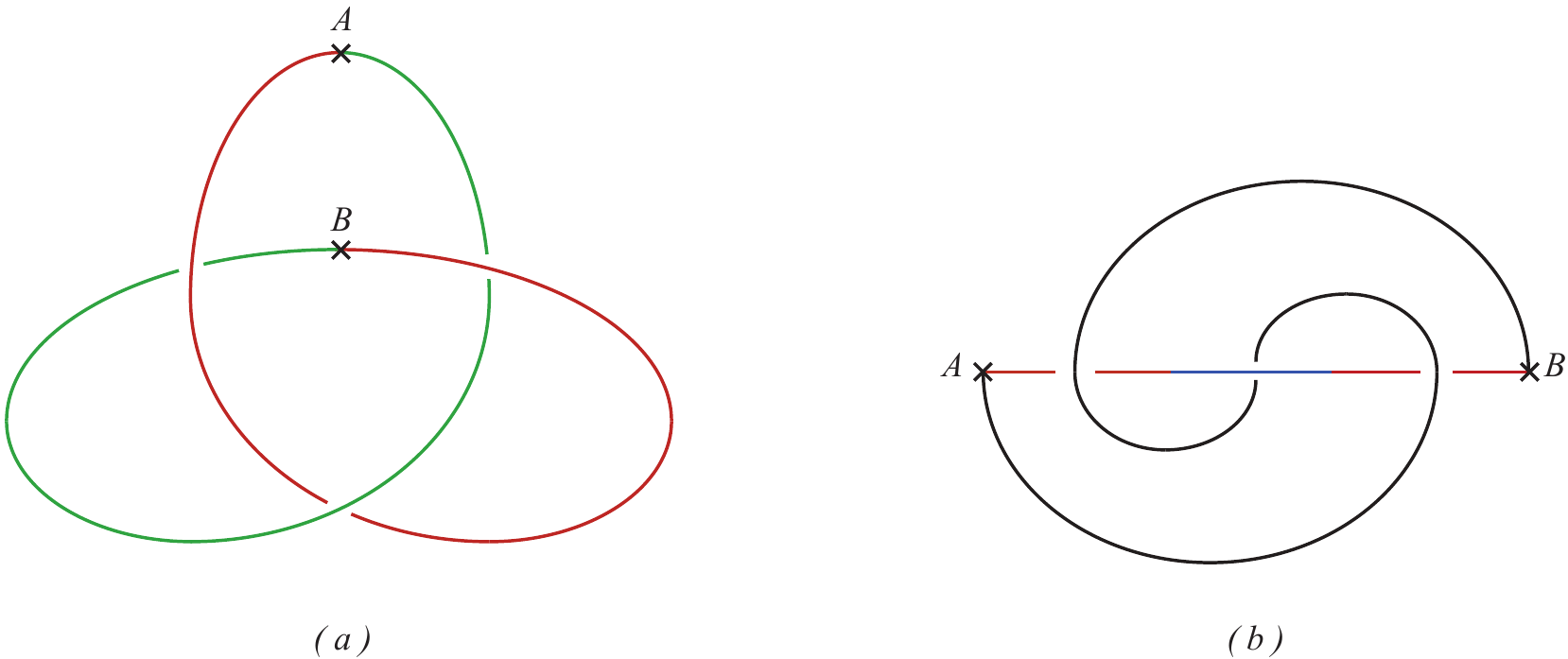}{8cm}{Two representation methods of a trefoil}

Later, Reidemeister considered a variation of such a representation and proved that each knot can be embedded into a plane so that it can be decomposed into one straight line segment and one non-self-intersecting string. Further, he left the larger string (that he called Faden $\mathcal F$) in the plane and moved the straight line in ${\mathbb R}^3$ forming an arcade $\mathcal A$ with arcs alternating on the upper and the lower side of the Faden plane \cite{Reidemeister}. He called such a representation {\it ''Arkaden Faden Lage``} (the arcade-strand-position), AFL for short. Figure \ref{Trefoil-1}(b) shows an AFL representation of a trefoil.

Each arcade is denoted either by $T$ (upper side of the plane) or by $S$ (lower side of a plane). Each such arcade has an index -- enumerated in the direction of the orientation of $\mathcal A$. Furthermore, if the orientation of the Faden is "UP", we write $S_i$ (resp. $T_i$) and if it is "DOWN", we write $S_i^{-1}$ (resp. $T_i^{-1}$). Thus, moving in the direction of faden $\mathcal F$, we get a unique word describing the AFL of a given knot.
For example, the word for a trefoil shown in fig. \ref{Trefoil-1}(b) is $S_1^{-1}T_2S_3^{-1}$. Note that the word for its opposite trefoil would be $T_1^{-1}S_2T_3^{-1}$.

Based on the above idea, Hotz \cite{Hotz2,Hotz1} proved that each knot builds a unique set of so-called minimal words over a given alphabet.
Since there are different AFL representations of one knot, each knot builds a unique set of minimal words. If two knots $K_1$ and $K_2$ build two sets of minimal words $S_1$ and $S_2$, then $K_1$ and $K_2$ are equivalent if and only if $S_1=S_2$ (or, alternatively, they have at least one common minimal word --- common AFL representation).
Using these ideas, he presented an $O(n^2\cdot2^{\frac{n}{3}})$ time bounded algorithm for a knot problem in \cite{HotzKnots} (here, $n$ is the number of crossings in the AFL representation of the given knot).
An efficient algorithm to get an AFL representation of a knot from its projection is also due to G. Hotz.

Around 1989, V. Vassiliev \cite{Vass1,Vass2} and M. Goussarov \cite{Gous} 
independently introduced the notion of so-called {\it finite type invariants} thus providing a radically new way of looking at knots. Vassiliev's approach is based on the study of discriminants in the (infinite-dimensional) spaces of smooth immersions from one manifold into another. The finite type invariants are also cited as Vassiliev invariants and are at least as strong as all known polynomial knot invariants: Alexander, Jones, Kauffman, and HOMFLY polynomials. This means that if two knots $\mathcal K_1$  and $\mathcal K_2$ can be distinguished by such a polynomial, then there is a Vassiliev invariant that takes different values for them.

One of the most powerful tools to compute Vassiliev invariants is the Kontsevich integral invented by Maxim Kontsevich in 1993. In fact, it is a far-reaching generalization of the Gauss integral for the linking number. Roughly speaking, given a knot $\mathcal K$ embedded in $\mathbb R^3$, it computes an appropriate rational number (defined by $\mathcal K$) for any chord diagram (to be defined later). So, it defines the infinite series in the algebra of chord diagrams that is supposed to be unique for each isotopy class of knots.

Further, in the late 1990s, Vassiliev \cite{Vassiliev} introduced the holonomic parametrization of knots by considering a periodic function $f$, where $(-f(x),f'(x),-f''(x))$ gives the parametrization of the knot in Cartesian coordinates. He showed that, for each knot $\mathcal K$, there exists a knot $\mathcal K'$ equivalent to $\mathcal K$ with a holonomic parametrization, but no method to find such a function was known.

More precisely, Vassiliev proved that any knot class (topological isotopy class of knots) has a holonomic representative and also that there exists a natural isomorphism from finite type invariants of topological knots to finite type invariants of holonomic knots.

Birman and Wrinkle \cite{Birman} showed that two holonomic knots which are topologically
isotopic are in fact holonomically isotopic. From a combinatorial point
of view this means that the holonomic isotopy classification of holonomic knots
is identical to the isotopy classification of their diagrams (an {\it isotopy of a knot
diagram} is defined to be a sequence of planar isotopies and Reidemeister moves). Therefore, many algorithms on knots (such as the knot isotopy algorithm in  \cite{HotzKnots}) could be improved by considering only holonomic knots.

\section{Basic Definitions and
preliminary remarks}

\subsection{AFL Representation of knots}

Consider an AFL representation of a trefoil shown in fig. \ref{Trefoil1}(a). The arcade AD is divided in three parts AB, BC and CD, where AB and CD build an arcade  under the larger curve (called Faden or $\mathcal{F}$) in ${\mathbb R}^3$. On the contrary, BC builds an arcade over $\mathcal{F}$. It is the minimal AFL representation of a trefoil, since it has no 'redundant' parts as shown in fig. \ref{Trefoil1}(b). Having such a non-minimal AFL representation of a knot, we reduce it by appropriate Reidemeister moves.

\Figure{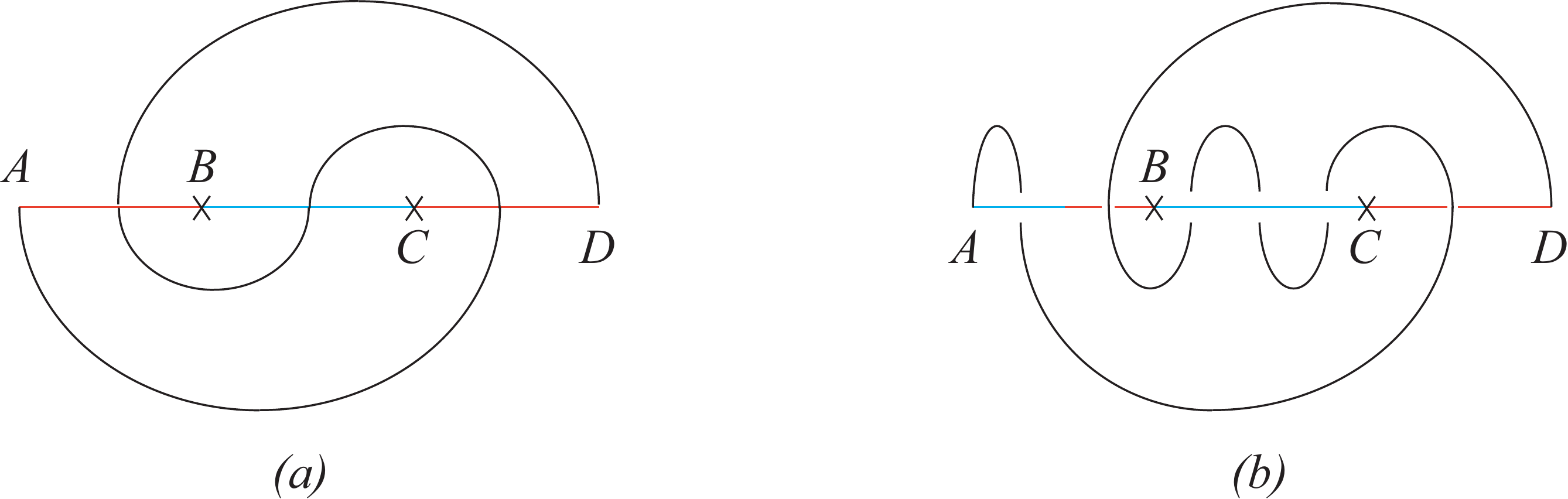}{8cm}{Minimal AFL (a) and AFL with redundant parts (b)}

Each AFL $F$ defines a word $\sigma_F$ over an infinite alphabet $\Sigma=\{T_i^\epsilon,S_i^\epsilon:i\in{\mathbb N}, \epsilon\in\{1, -1\}Ê\}$. Each part of the arcade is enumerated from 1 to $n$ in ascending order in the direction of its orientation. To each $i$-th part corresponds a variable $T_i$ or $S_i$, depending on the position of the given part of the arcade: if it builds an arcade under $\mathcal{F}$, we call it $S_i$, else $T_i$. In the example shown in fig. \ref{Trefoil1}(a), we have the variables $S_1,T_2$ and $S_3$. Depending on the orientation of $\mathcal{F}$, we write $S_i^{-1}$ (or $T_i^{-1}$ respectively) if the projection of $\mathcal{F}$ crosses the arcade top-down, and $S_i^{1}$ ($T_i^{1}$ respectively) else.
In our example, we have $S_1^{-1},T_2^1$ and $S_3^{-1}$. Defining the word for a given knot, we must arrange these variables in the same order as the projection of  $\mathcal{F}$ crosses the projection of the arcade. For the trefoil in the example above we again get ''$S_1^{-1}\;T_2^1\;S_3^{-1}$``.

According to these rules, we get for the trefoil representation in fig. \ref{Trefoil1}(b):  ''$S_2^{-1}\;T_3^1\;T_3^{-1}\;T_3^1\;S_4^{-1}\;S_1^1$``. 

\subsection{Holonomic representation of knots}

This section is based on \cite{Birman}.

Let $f:\mathbb R\rightarrow\mathbb R$ be a $C^{\infty}$ periodic function with period $2\pi$. Following Vassiliev
\cite{Vassiliev}, use $f$ to define a map $\tilde{f}:S^1\rightarrow{\mathbb R}^3$ by setting $\tilde f(t) = (-f(t),f'(t),-f''(t))$. Let $\phi$ be the restriction of  $f$ to the first two coordinates. We call $\phi$ the projection
of $K = \tilde f(S^1)$ (onto the xy plane).

A simple example is
obtained by taking $f(t) = \cos(t)$, giving the unknot. Another example is given in figure  \ref{TrefoilNew}.

\Figure{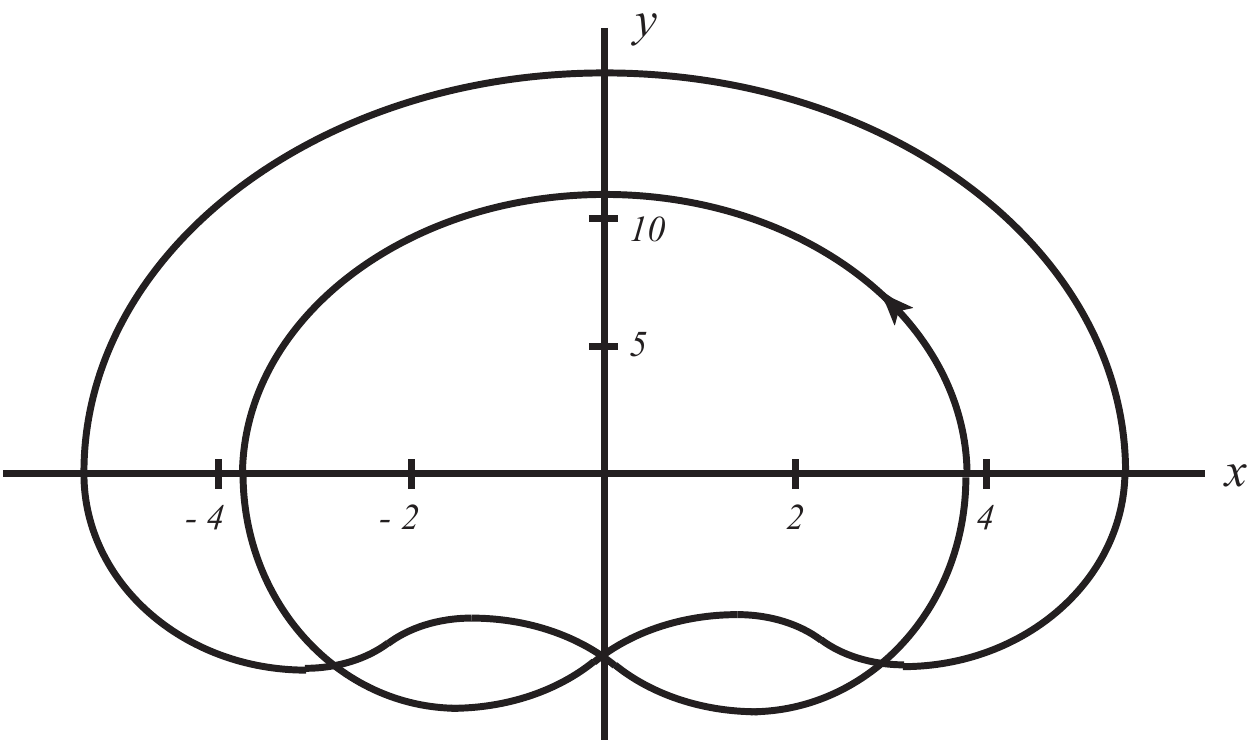}{3cm}{The function $f(t) = \sin(t) + 4\sin(2t) + \sin(4t)$ determines a holonomic trefoil}

Two important things have to be noted:

\begin{enumerate}
\item The orientation of a holonomic knot is always counter-clockwise;
\item Consult Figure \ref{UpDown}, which
shows four little arcs in the projection of a typical $K = \tilde f(S1)$ onto the xy plane. The four strands are labeled 1,2,3,4. First consider strands 1 and 2. Both are necessarily
oriented in the direction of decreasing $x$ because they lie in the half-space defined by $f'(t) > 0$. Since $f'$ is decreasing on strand 1, it follows that $f''$ is negative on strand 1,
so $-f''$ is positive, so strand 1 lies above the $xy$ plane. Since $f'$ is increasing on strand
2, it also follows that strand 2 lies below the $xy$ plane. Thus the crossing associated
to the double point in the projection must be negative, as in the top sketch in Figure \ref{UpDown} (b), and in fact the same will be true for every crossing in the upper half-plane. For
the same reasons, the projected image of every crossing in the lower half of the $xy$
plane must come from a positive crossing in 3-space.
\end{enumerate}

\Figure{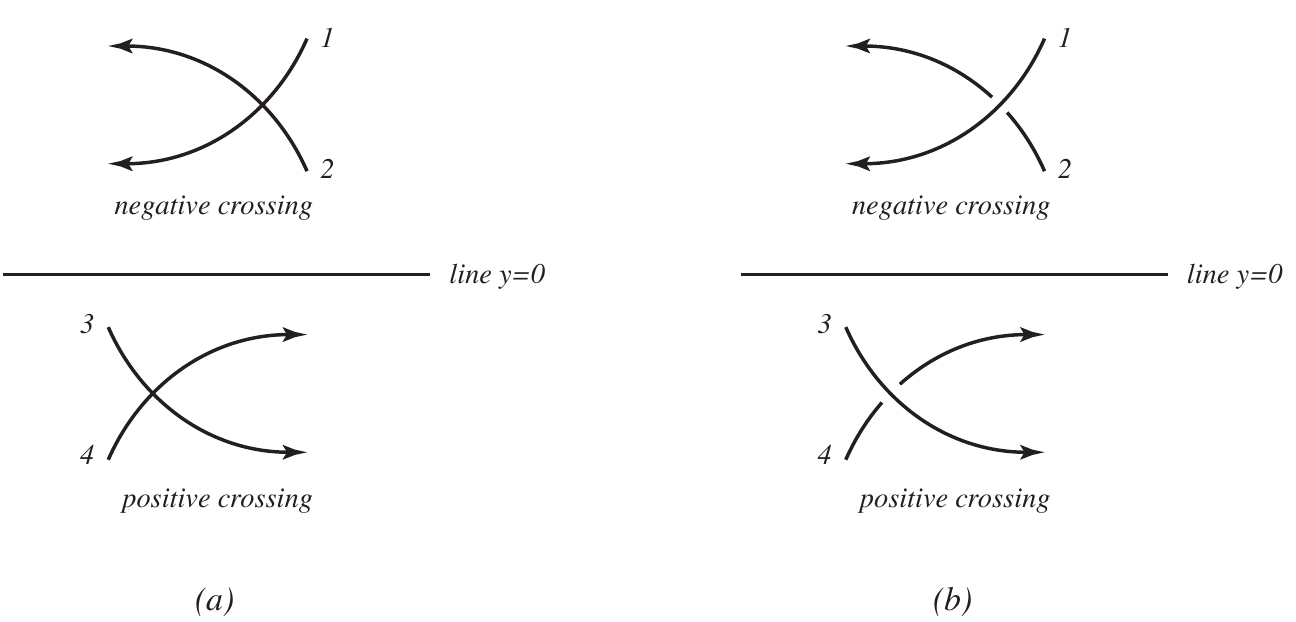}{8cm}{Determining the sign of a crossing: Figure (a) shows the projected images onto the xy plane; Figures (b) and (c) show their lifts to 3-space.}

\subsection{The Kontsevich Integral for Knots}

This section is based on  \cite{ChmuDu}.

Consider a trefoil knot $\mathcal T$ in ${\mathbb R}^3$ as shown in fig.  \ref{3D_TrefoilExtr}.

\Figure{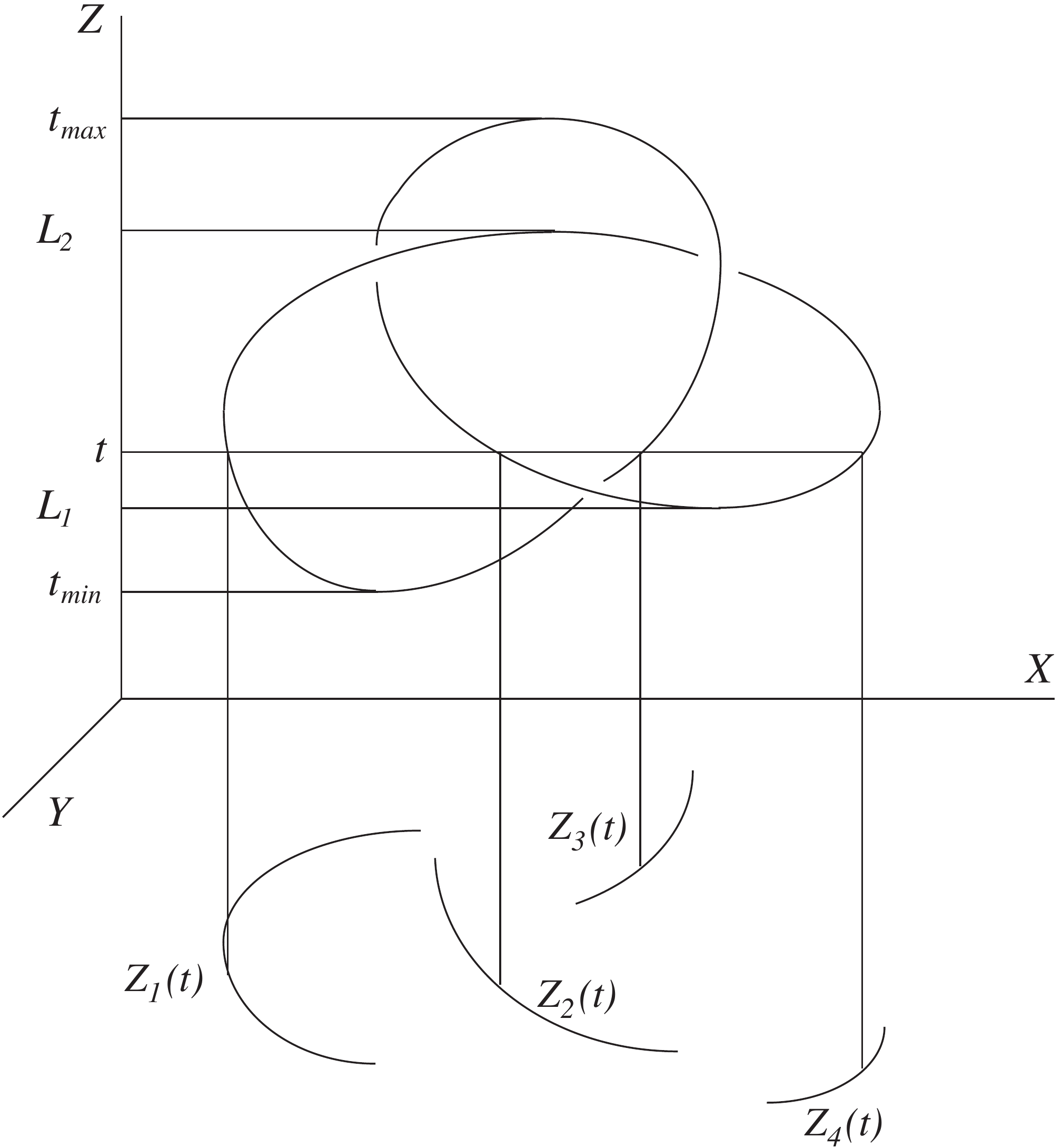}{6cm}{Trefoil in ${\mathbb R}^3$}

The three-dimensional space can be represented as a direct product of a complex plane
${\mathbb C}$ with coordinate $z$ and a real line ${\mathbb R}$ with coordinate $t$.

Note that the Kontsevich integral is defined for Morse knots, i. e. knots $\mathcal K$ embedded in ${\mathbb R}^3 ={\mathbb C}_z \times {\mathbb R}_t$ in such a way that the coordinate $t$ restricted to $\mathcal K$ has only nondegenerate (quadratic) critical points.

Define a set 
$\{Z_1,...,Z_l:Z_i=(a_i,b_i,t)\in\mathcal T\}$
of the points of $\mathcal T$ with the projections $Z_i(t)$ on $\mathbb C$ for each coordinate $t$ (this set can be empty).
We consider each $Z_i(t)$ as a continuous function $Z_i:T\subset{\mathbb R}\rightarrow{\mathbb C}$. The ${\mathbb R}$ axis is divided into segments by the extremal points of the given knot, in our example the critical points $t_{min}$, $L_1$, $L_2$ and $t_{max}$ as shown in fig. \ref{3D_TrefoilExtr}.
Each function $Z_i(t)$ is defined in one of these segments.

Furthermore, we consider each knot $\mathcal K$ as a continous function from a circle $\mathcal R$ into ${\mathbb R}^3$.
Defining direction in $\mathcal R$ we also define a direction on $\mathcal K$.
Choosing two pairs of points in $\mathcal K$, i.e. $(Z_1, Z_2)$ and $(P_1,P_3)$ as shown in fig.  \ref{3D_Trefoil2}, and connecting the corresponding points on the circle, we get a so-called chord diagram. Another pairs, i.e. $(Z_1, Z_3)$ and $(P_1,P_2)$ define another chord diagram.

\Figure{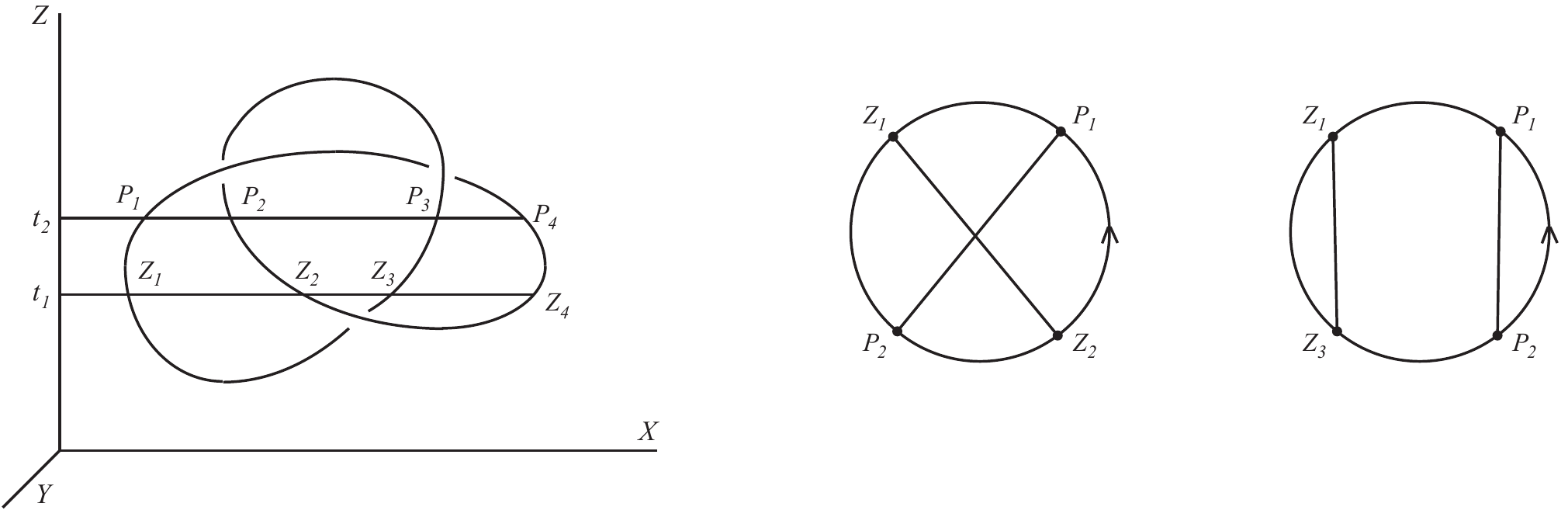}{8cm}{2-chord diagrams}

Similarly, $n$ pairs of points on $\mathcal K$ define an $n$-chord diagram. Obviously, two different sets with $n$ pairs of points on $\mathcal K$ can define the same chord diagram. Fig. \ref{CordDiagr} shows the examples of $3, 4, 5$ and $6$-chord diagrams.

\Figure{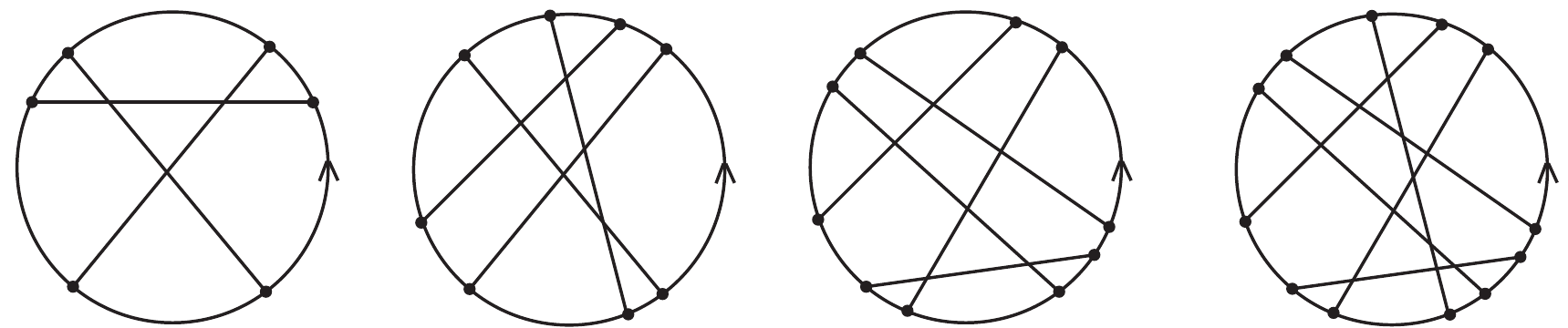}{5cm}{Examples of $3, 4, 5$ and $6$-chord diagrams}

Now consider two sets of points $\{Z_1,Z_2,Z_3,Z_4\}$ and $\{P_1,P_2,$ $P_3,P_4\}$. The points within each set have the same  coordinates $t_1$ and $t_2$ respectively as shown in fig. \ref{3D_Trefoil3}.

\Figure{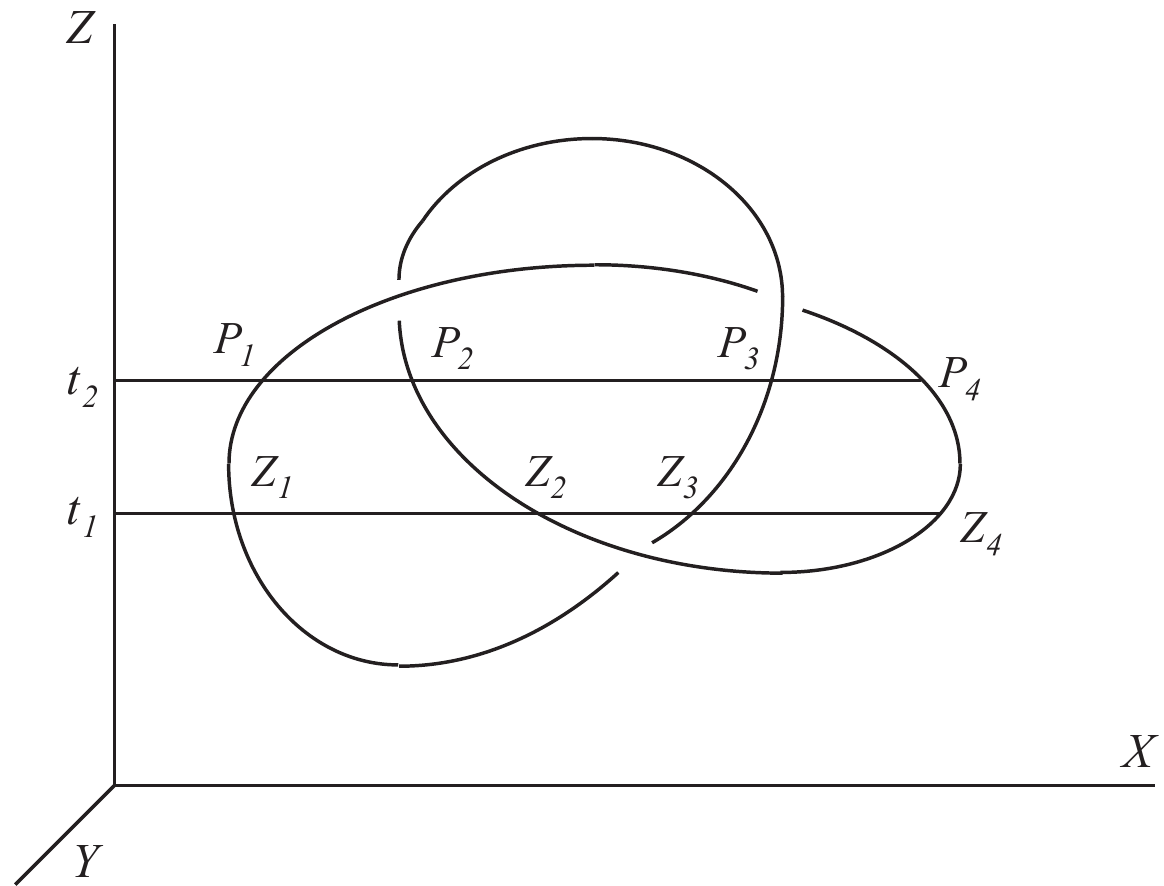}{5cm}{Two sets of points with same $Z$ coordinates}

From each set, we choose one pair of points and define a corresponding chord diagram. There are 36 different possibilities for choosing all the possible pairings from each set, thus defining the 2-chord diagrams shown above. If we have not two but $m$ such sets of points, we can define an $m$-chord diagram by choosing one pair of points from each set.

The Kontsevich Integral is calculated by the following formula:

$
\begin{array} {ll}
&\sum_{m=1}^\infty\frac{1}{(2\pi\cdot i)^m}\cdot\\
&\int\limits_{{t_{min}<t_1<\cdots<t_m<t_{max}}\atop{t_i\; noncritical}}\sum_{(z_i,z'_i)\in P}(-1)^{\#P\downarrow}D_P\bigwedge\limits_{j=1}^m\frac{d(z_j-z'_j)}{z_j-z'_j}.
\end{array}
$

The integration domain is the $m-$dimensional simplex $t_{min}<t_1<\cdots<t_m<t_{max}$ divided by the critical values into a certain number of connected components. For example, for the following embedding of a trefoil and $m=2$  the corresponding integration domain has six connected components and looks as shown in fig. \ref{Domain}.

\Figure{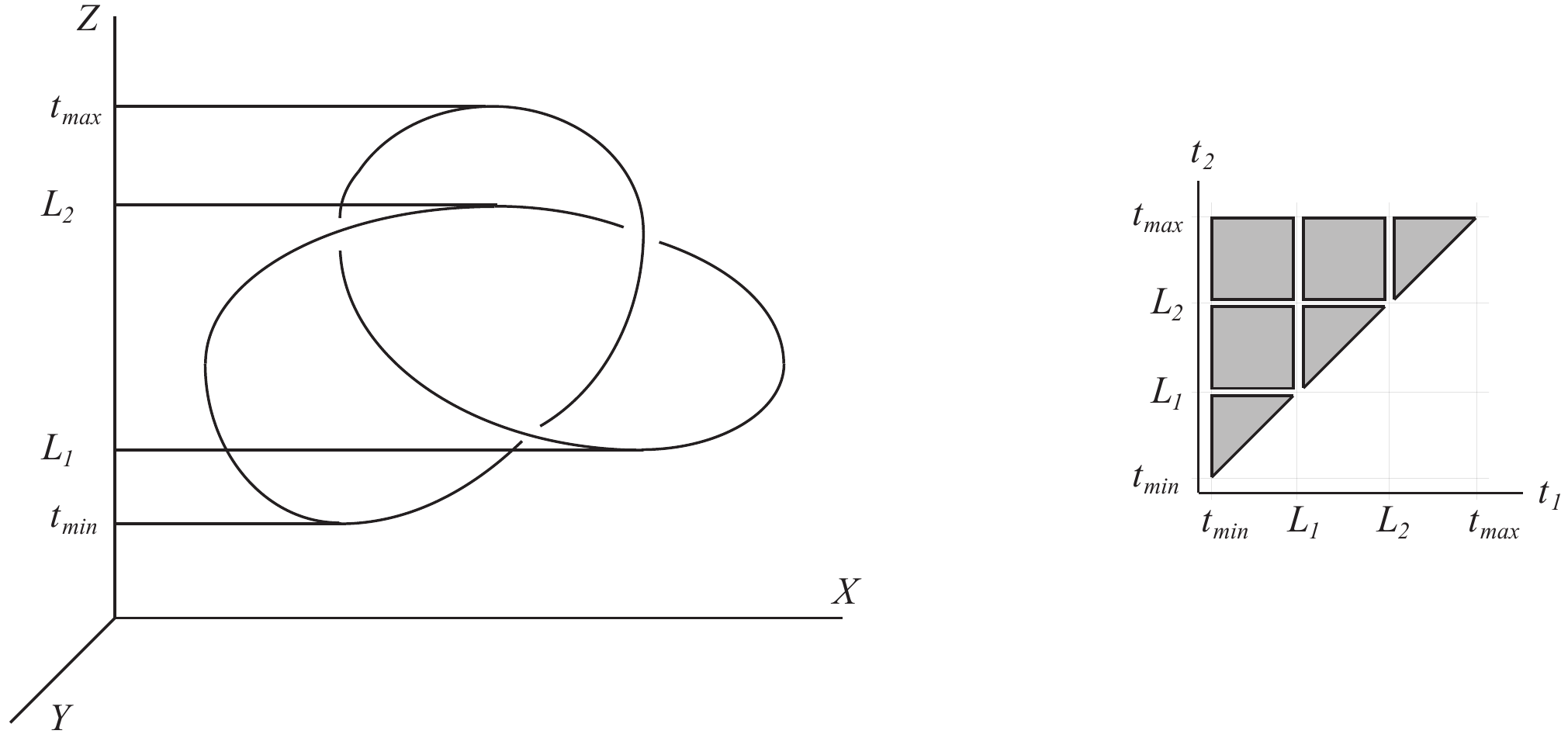}{8cm}{Connected components of the trefoil}

The number of summands in the integrand is constant in each connected component of the integration domain, but can be different for different components. In each plane $\{t=t_j\}\in {\mathbb R}^3$ choose an unordered pair of distinct points $(z_j, t_j)$ and $(z'_j, t'_j)$ on $\mathcal K$, so that $z_j(t_j)$ and $z'_j(t_j)$ are continuous functions. We denote by $P=\{(z_j,z'_j)\}$ the set of such pairs for $j=\overline{1;m}$. The integrand is the summand over all choices of $P$.
In the example above for the component $\{t_{min}<t_1<t_2<L_1\}$ we have only one possible pair of points on the levels $\{t=t_1\}$ and $\{t=t_2\}$. Therefore, the sum over $P$ for this component consists of only one summand. 
Unlike this, in the component $\{t_{min}<t_1<L_1<t_2<t_{max}\}$ we still have only one possibility for the level $\{t=t_1\}$, but the plane $\{t=t_2\}$ intersects the trefoil knot $\mathcal K$ in four points. So we have $\left(4\atop2\right)=6$ possible pairs $(z_2, z'_2)$ and the total number of summands is six.
On the other hand, in the component $L_1<t_1< t_2<L_2\}$ each of the plains $\{t=t_1\}$ and $\{t=t_2\}$ intersect $\mathcal K$ in four points building six possible pairs each and 36 summands. 

For a pairing $P$, the symbol '$\#P\downarrow$' denotes the number of points $(z_j, t_j)$ or $(z'_j, t_j)$ in $P$ where the coordinate $t$ decreases along the orientation of $\mathcal K$.

By fixing a pairing $P$, we define an appropriate chord diagram $D_P$ with $m$ chords as described earlyer in this work. 

Figure \ref{Summands} shows one of the possible pairings for each connected component in our example as well as the corresponding chord diagram with the sign $(-1)^{\#P\downarrow}$ and the number of summands of the integrand.

\Figure{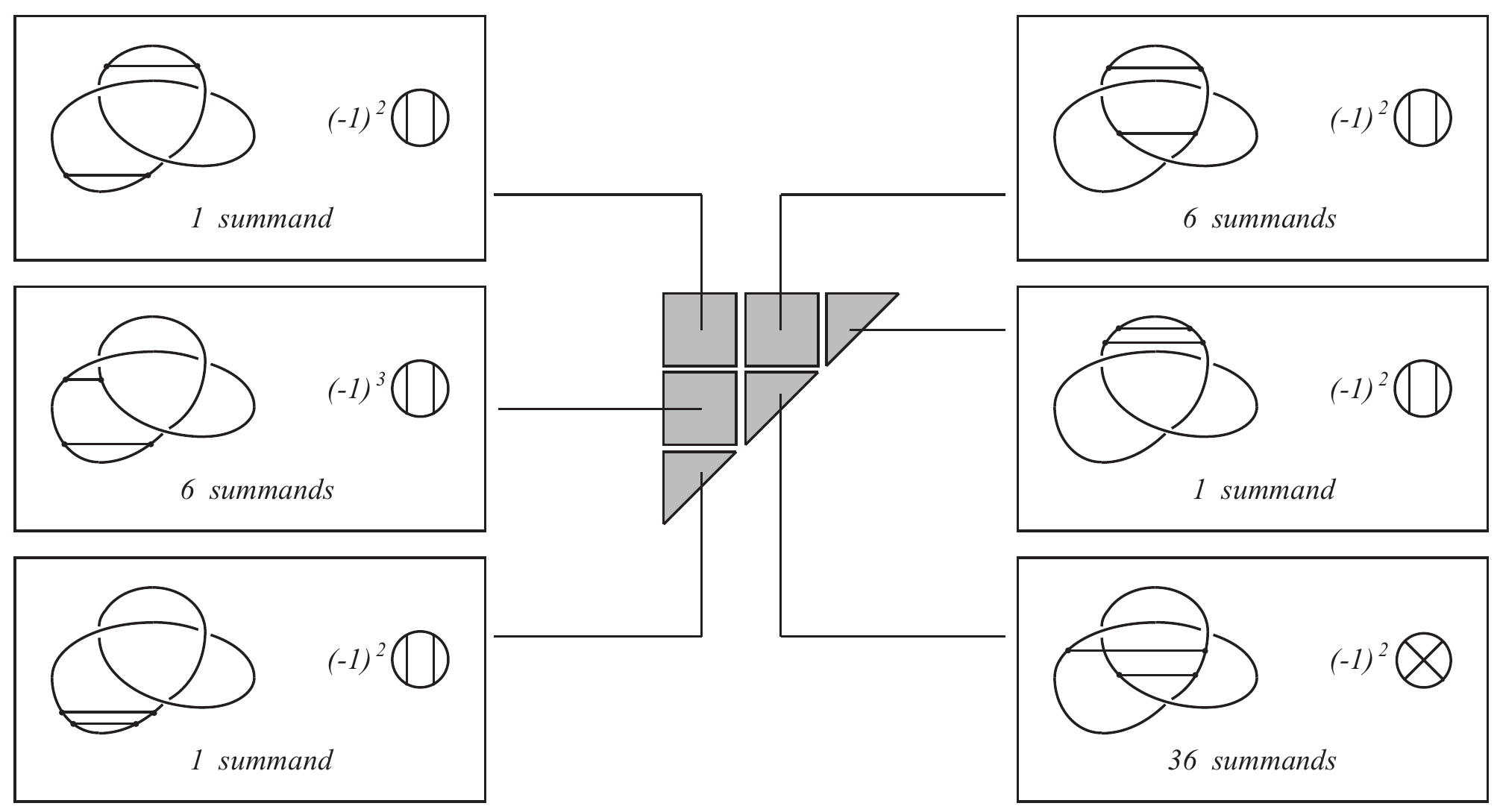}{8cm}{Examples of pairings in connected components}

Over each connected component, $z_j$ and $z'_j$ are smooth functions in $t_j$. By $\bigwedge\limits_{j=1}^m\frac{d(z_j-z'_j)}{z_j-z'_j}$ we mean the pullback of this form to the integration domain of variables $t_1,...,t_m$. The integration domain is considered with the positive orientation of the space ${\mathbb R}^m$ defined by the natural order of the coordinates $t_1,...,t_m$.

Roughly speaking, given a fixed knot and any chord diagram, the Kontsevich integral gives a {\it rational} number for this specific chord diagram. Thus, given an infinite sequence of chord diagrams, we can define a corresponding infinite sequence of rational numbers that is supposed to be unique for each isotopy knot class (it is supposed that two knots have the same rational numbers for {\it each} chord diagram if and only if they are isotopic).

\section{Using AFL in the
 Computation of the Holonomic Representation of Knots}

Consider the four steps of the construction of the faden $\mathcal  F$ of the trefoil AFL shown in fig. \ref{TrefoilConstr1}. In each step, we construct a corresponding curve $C_i, i=1,...,4$.

\Figure{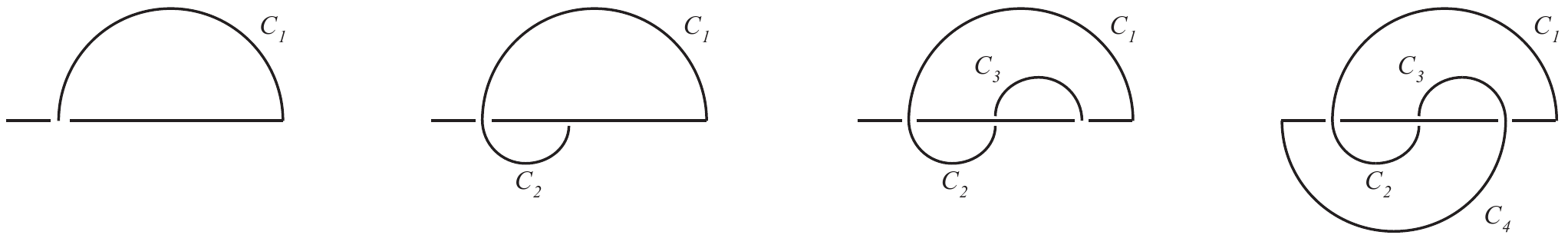}{8cm}{Construction steps of a trefoil AFL}

Obviously, the curves $C_1$ and $C_2$ can be easily described holonomically, but the curves $C_3$ (from point $A$ to point $B$) and $C_4$ (from point $B$ to point $E$) are not holonomic because of their clockwise orientation (see fig. \ref{TrefoilHolonomic1}(a)). Thus we have to make some changes in the AFL representation.

\Figure{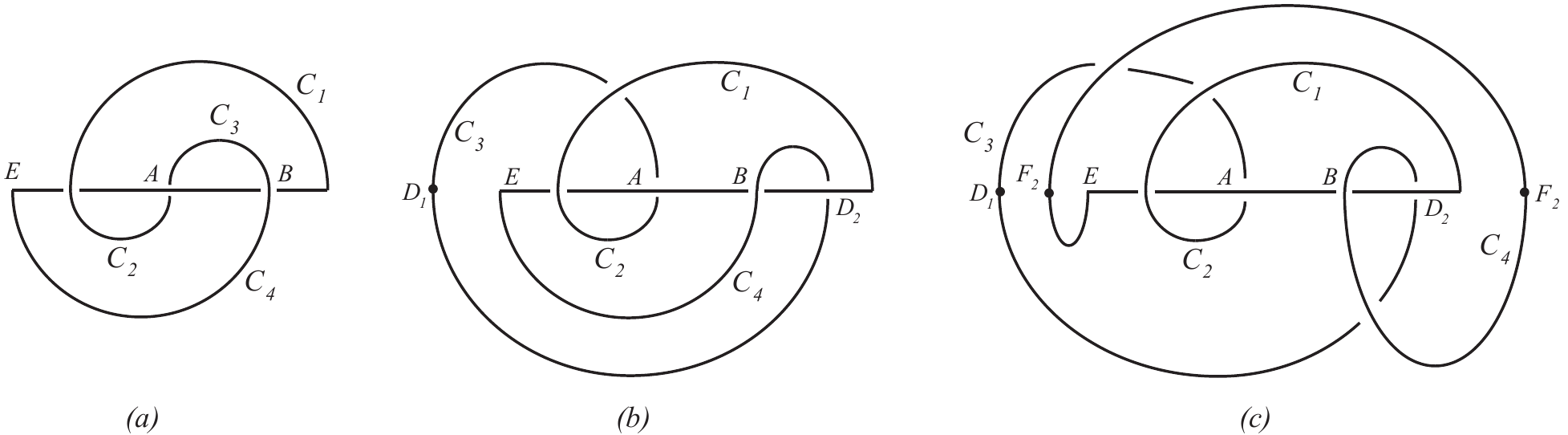}{8cm}{Construction steps of the holonomic AFL}

First of all, we draw the curve $C_3$ from point $A$ to point $B$ passing the new point $D$ as shown in fig. \ref{TrefoilHolonomic1}(b). The main idea is to draw a counter-clockwise oriented curve preserving the topological structure of the original construction (it could be restored by Reidemeister moves). The curve $C_4$ is also rearranged by a similar idea (fig. \ref{TrefoilHolonomic1}(c)).

\Figure{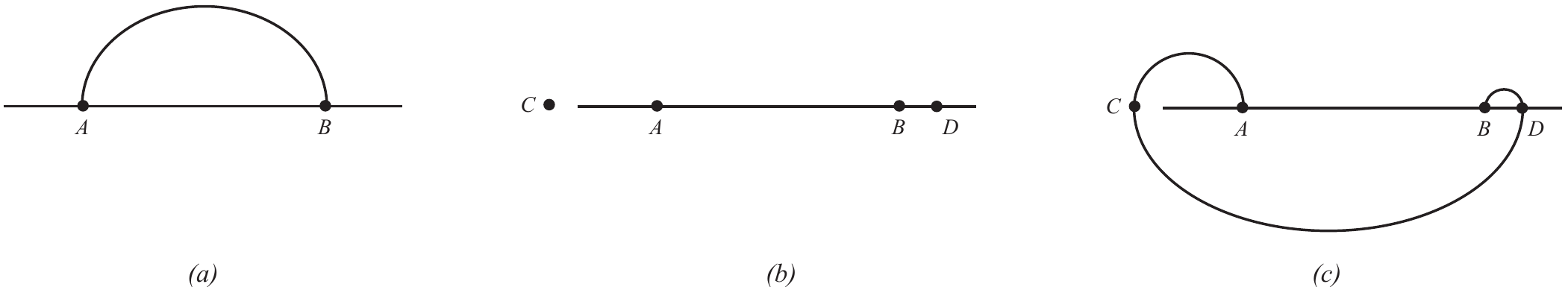}{8cm}{Changing the orientation of a curve}

In general, if we have a clockwise oriented curve from point $A$ to point $B$ as shown in fig. \ref{HolonomicConstruction-1}(a), we first define two additional points $C$ (that is a leftmost point on the prolongiation of the arcade line) and $D$ (that is the point in the very right neighborhood of $B$).

Similar to this, we can change the orientation of the curve shown in fig. \ref{HolonomicConstruction1-1}. The only difference here is that we define the {\it rightmost} point on the prolongiation of the arcade line $C$ and a point $D$ in the {\it very left} neighborhood of $B$.

\Figure{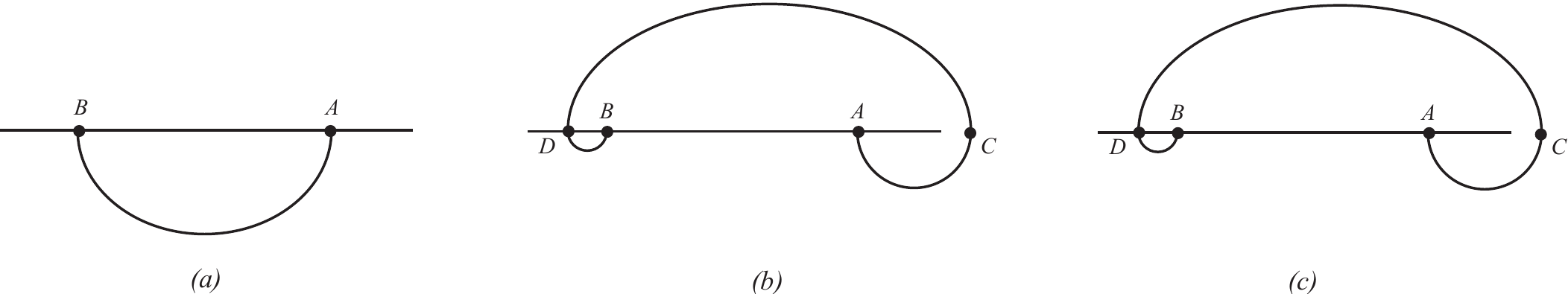}{8cm}{Changing the orientation of a curve}

This construction does not violate the topology of the original $AFL$. As we can see from fig.  \ref{HolonomicTopology3-1}, if the curve with the clockwise orientation (from point $A$ to point $B$) is in the upper half plane from the arcade line, the reconfigured curve lies completely {\it under} all other curves (fig. \ref{HolonomicTopology3-1}(a), (b), (e) and (f)); if the curve with the counter-clockwise orientation (from point $A$ to point $B$) is in the lower half plane from the arcade line, the reconfigured curve lies completely {\it over} all other curves (fig. \ref{HolonomicTopology3-1}(c), (d), (g) and (h));

\Figure{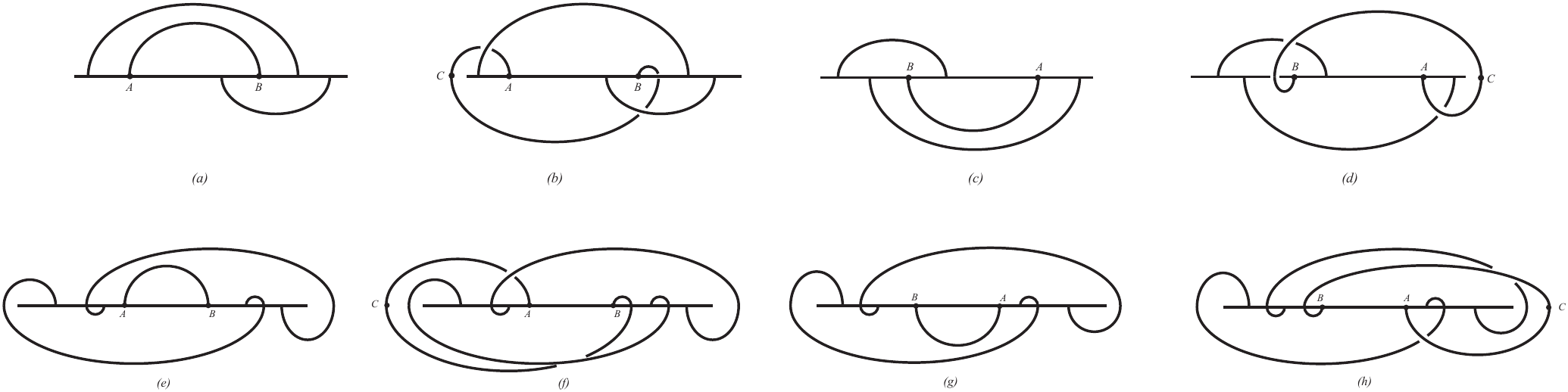}{8cm}{The relative position of a reconfigured curve}

Note that all curves in the diagrams above except the curve from $A$ to $B$ have counter-clockwise orientation.

In the construction process, it is important to reconfigure the curves in a proper order. Consider the situation in fig. \ref{ConstructionTopology-1}.

\Figure{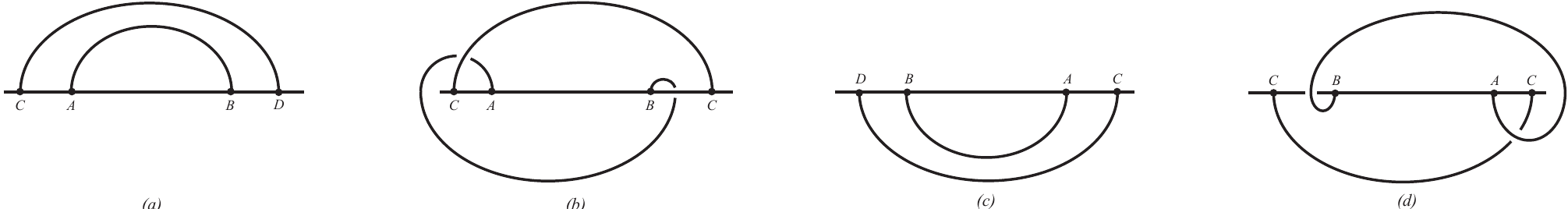}{8cm}{The improper reconfiguration order}

Here we have two nested clockwise-oriented curves from point $A$ to point $B$ and from point $C$ to point $D$ (fig. \ref{ConstructionTopology-1} (a) resp. (c)). If we first reconfigure the inner curve we get the situations shown in fig. \ref{ConstructionTopology-1} (b) and (d). In both cases we can no longer rearrange the curves from point $C$ to point $D$ in a way described above because these curves lay {\it under} the rearranged inner curves. The solution is to reconfigure the {\it inner} curve {\it after} the reconfiguration of the outer curve.

Till now, we have treated the crossings of the {\it Faden} curves with the {\it Arcade} line as black dots. In reality, the {\it Arcade} is passing over or under the {\it Faden}. Assuming that the direction of the {\it Arcade} line is from left to right, some non-holonomic  situations can occur as shown in fig.  \ref{HolonomicCrossings-2}.

\Figure{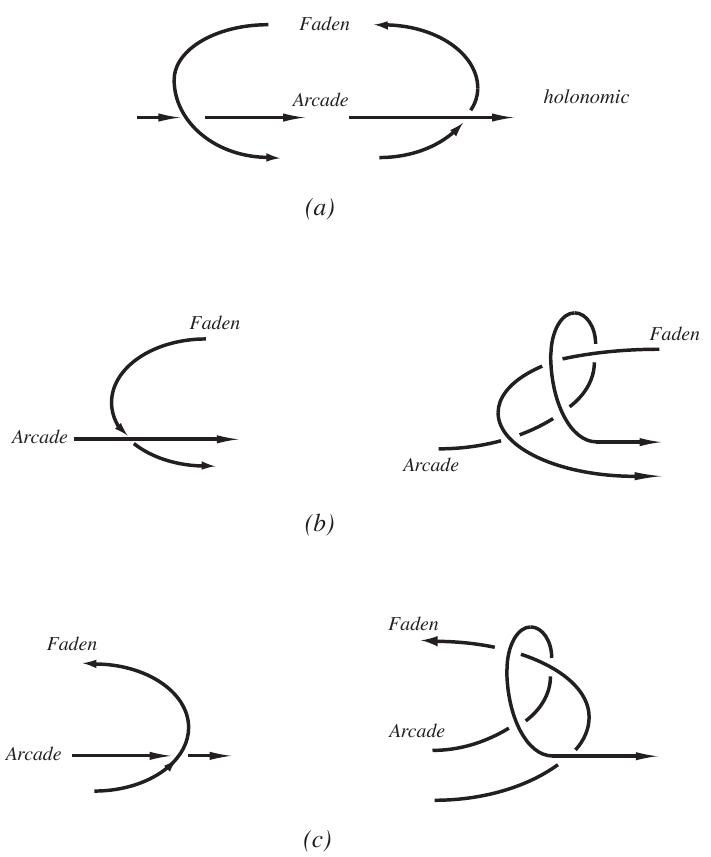}{8cm}{Rearranging non-holonomic crossings}

Fig.  \ref{HolonomicCrossings-2}(a) shows holonomic crossings, but the crossings shown on the left side of the fig. \ref{HolonomicCrossings-2} (b) and (c) are non-holonomic and should be rearranged as shown on the right sides of the appropriate figures.

Thus we get the following algorithm to compute the holonomic function for a given AFL:

{\bf Input:} AFL representation of a knot;\\ \\
1. For each non-holonomic curve define additional points and describe the holonomic curve connecting these points by a holonomic function;\\
2. For each non-holonomic crossing of the arcade $\mathcal A$ and the rearranged curves define additional points and describe the holonomic curve connecting these points by a holonomic function.

Based on this algorithm, we get for the holonomic representation of a trefoil diagram shown in fig.  \ref{TrefoilHolonomic1-1}(a).
Note that in most cases, these representations are not optimal, so one can apply some optimization algorithms to get an optimal or near-optimal holonomic representation as shown in 
fig. \ref{TrefoilHolonomic1-1}{\it(a) - (d)}.

\Figure{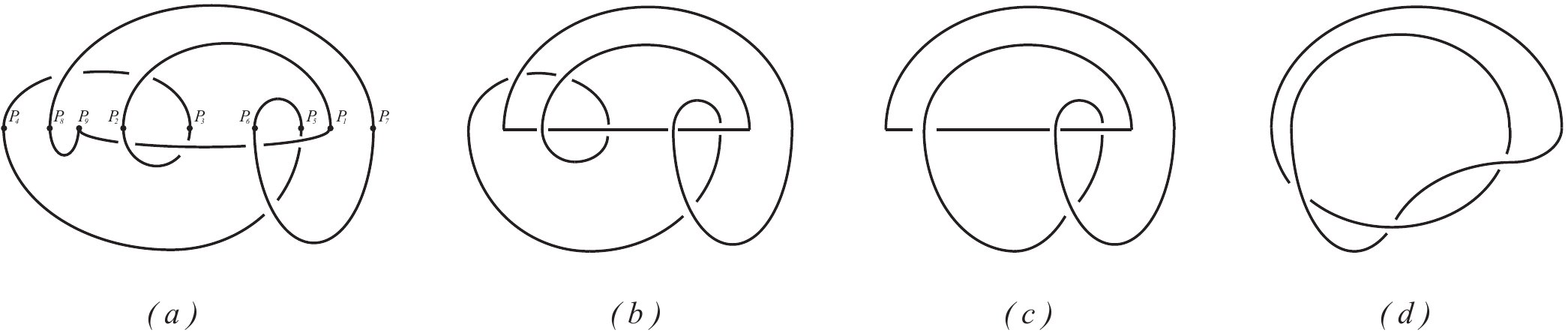}{8cm}{The holonomic representation of a trefoil}

Tu sum up the results, each knot described as an AFL can be rearranged as a holonomic knot using the ideas above. 
Due to this, after the rearrangement process, we get a set of points $P=\{P_1,...,P_k\}$, where the points $(P_i,P_j), j=i+1, 1\leq i<j\leq k$ as well as $(P_k,P_1)$ must be connected by a curve. These counter-clockwise oriented curves can be easily described by a holonomic (trigonometric) function. So, for each curve connecting the points $(P_i,P_j)$ we get a function $f_i$. These $k$ functions can be combined to one holonomic representation of a knot using $f_i$ and a special type of a square wave function.
Since the rearrangement process always requires a limited number of steps for each non-holonomic curve or crossing, its computational complexity must be linear in the number of crossings of the AFL description of given knot.

\section{Further Possibilities: Using AFL in the\\
Computation of the Kontsevich Integral}

Consider the AFL representation of a trefoil knot in ${\mathbb R}^3$ so that it has the projection into the $XY$ plane as shown in fig. \ref{TrefoilAFL}(a).

\Figure{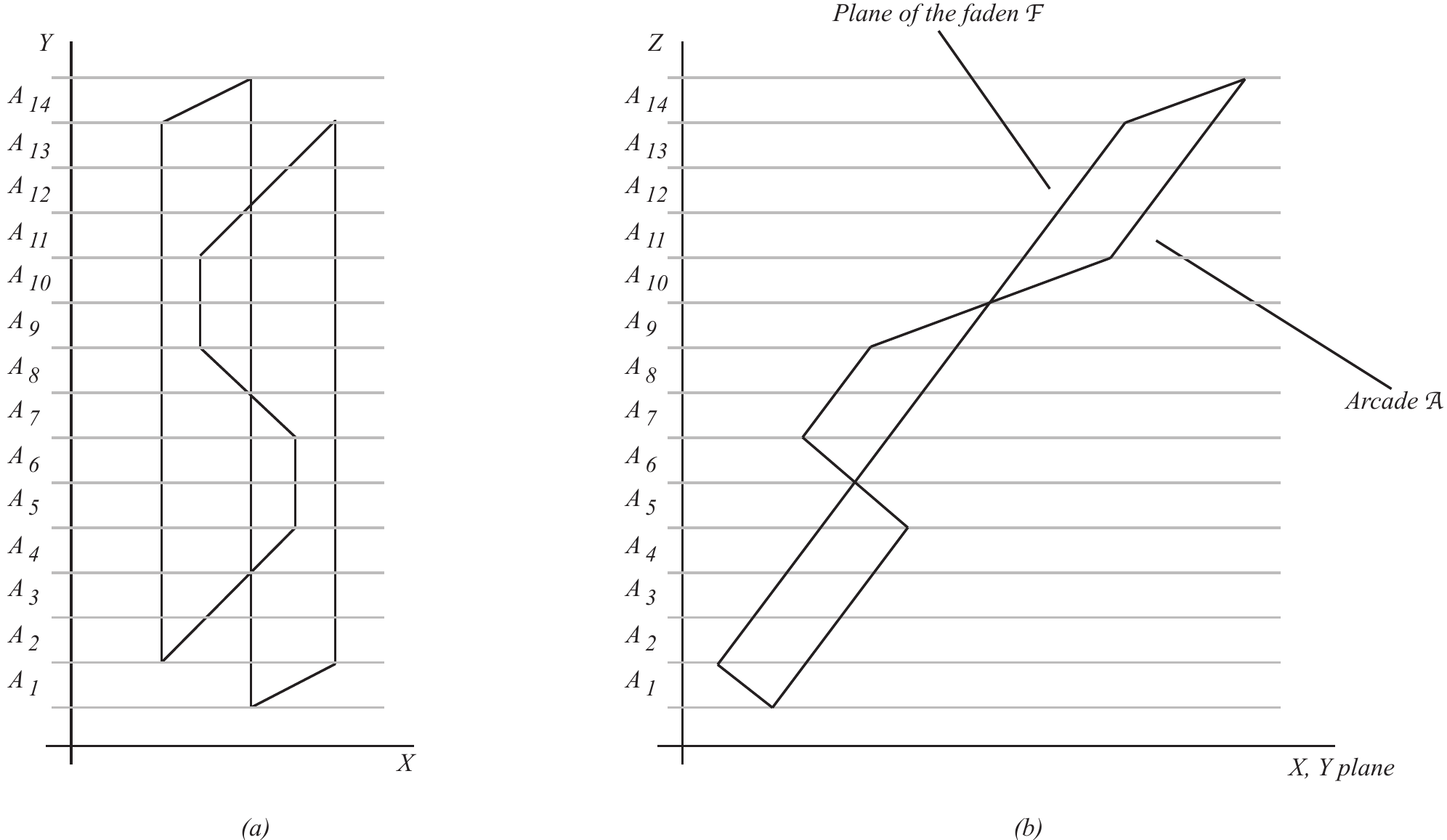}{8cm}{Trefoil projections}

The middle straight line parallel to the $Y$ axis is the projection of the arcade $\mathcal A$. The rest is the projection of the faden $\mathcal F$. Fig. \ref{TrefoilAFL}(b) shows the appropriate AFL from another perspective in ${\mathbb R}^3$ (its projection on the $XZ$ plane). Since $\mathcal F$ is placed in a plane, its projection is a straight line. Moreover, the arcade is placed in a plane parallel to the $XZ$ plane. 

Note that, in regions $A_1,...,A_4$ as well as in $A_{11},...,A_{14}$, the arcade is under the plane of $\mathcal F$. Unlike this, the arcade is above the plane of $\mathcal F$ in the regions $A_7,A_8$. In regions $A_5,A_6$ and $A_9,A_{10}$, the arcade is alternating from the lower to the upper side of the plane and vise versa. Due to this, if we cut the knot with the plane $t_1$ and $t_2$ parallel to $XY$, the projections of the points on $\mathcal F$ are on the line parallel to the $X$ axis. The projection of the point on the arcade $\mathcal A$ is either above or under the line described above, depending on the relative position of $\mathcal A$ to the plane of $\mathcal F$ in ${\mathbb R}^3$ (see fig. \ref{AFL_Points1}). The relative positions of the points in the projection in sections $A_6,...,A_9$ are equivalent to one another, so are the positions in the sections $A_1,...,A_5,A_{10},...,A_{14}$.

\Figure{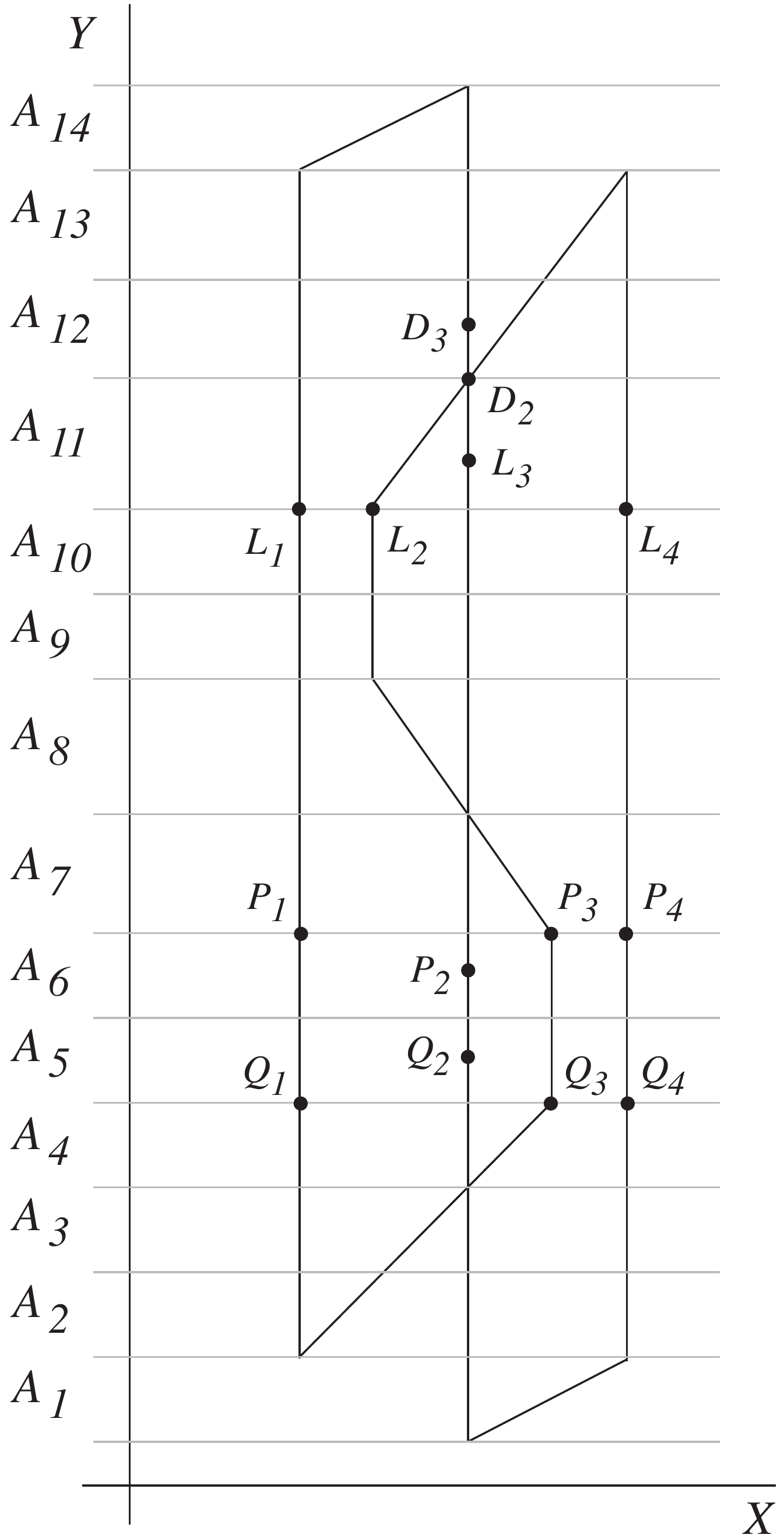}{4cm}{Examples of pairings in connected components}

For simplicity and w.l.o.g., we will use only the projection of the trefoil AFL as shown in fig. \ref{AFL_Points1} further in this work. Note that the higher the $Z$ coordinate of the cutting plane, the higher the $Y$ coordinates of the projections of the intersecting points.

Unlike the earlyer methods, the integration domain is the $m-$dimensional simplex $t_{min}<t_1<\cdots<t_m<t_{max}$ divided not only by the critical points but by 15 points shown in fig.  \ref{AFL_Points1} into a certain number of connected components.

The main advantage of this representation is that, choosing a pairing $P$ in each connected component, we {\it often} get a pair of points on parallel lines (e.g. $(P_1,P_4))$, thus the function $z_j(t)-z'_j(t)$ is constant and we get 
$\bigwedge\limits_{j=1}^m\frac{d(z_j-z'_j)}{z_j-z'_j}=0$ in $P$, so the number of summands will decrease dramatically. In our example, we get only 3 instead of 6 possible pairings $(P_1,P_2)$, $(P_2,P_3)$ and $(P_2,S_1)$. For each $m$, we get only $3^m$ summands instead of $6^m$ (in fact, this number even decreases because we get chord diagrams with isolated chords that can be ignored). In each area $A_i$ we get one {\it fixed} point for each pairing where the points are not lying on parallel lines.

Taking a 2-dimensional simplex $t_{min}<t_1<t_2<t_{max}$ where $t_1\in A_8$ and $t_2\in A_{11}$, we get the following non-zero pairings: $\{(P_1,P_2), (P_2,S_1), (P_2,P_3)\}$ and 
\\
$\{(P_1,P_2), (P_2,S_1), (P_2,P_3)\}$, where only $\{(P_2,S_1), (L_2,S_2)\}$, $\{(P_2,S_1), (L_2,L_1)\}$, $\{(P_2,P_3), (L_2,L_1)\}$, and \\ $\{(P_2,P_3), (L_2,S_2)\}$ build non-zero chord diagrams as shown in fig. \ref{ChordBuild}.

\Figure{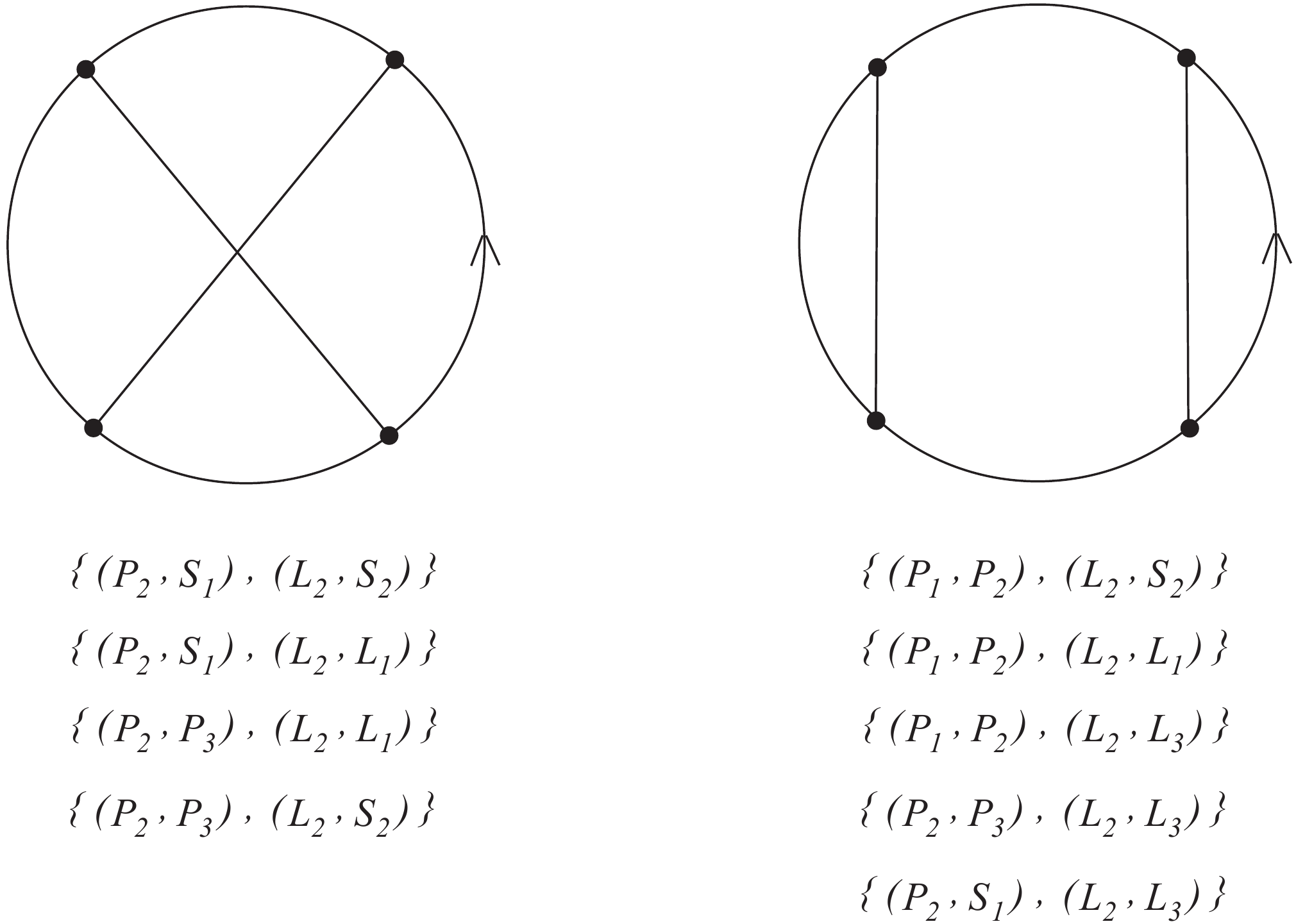}{4cm}{Point pairings and appropriate chord diagrams}

Note that we do not consider other pairings of points because they must contain at least one pair on parallel lines that can be ignored as discussed above.

Now consider a pair of points $(P_3,P_1)$ moving in $A_7$. Let their functions be $z_1(t)$ and $z'_1(t)$ respectively. Obviously, ${d\over dt}(z_1(t)-z'_1(t))>0$ in $A_7$. W.l.o.g. We can set $0<t<1$ and embed the trefoil in ${\mathbb R}^3$ so that $z_1(0)-z'_1(0)=3$ and, in general, $z_1(t)-z'_1(t)=3-t$ for the calculations in $A_7$ (for the moment $t=0$ we assume that the points are on the lower border of $A_8$ moving upwards to the upper border for the moment $t=1$). Since $(P_3,P_1)$ is always on a line parallel to the $X$ axis, $z_1(t)$ and $z'_1(t)$ always have equal imaginary part, so that $z_1(t)-z'_1(t)$ is a real number.
Hence we get

$$
\frac{\frac{d}{dt}(z_1(t)-z'_1(t))}{z_1(t)-z'_1(t)}=-\frac{1}{3-t}.
$$

We can embed the trefoil AFL in ${\mathbb R}^3$ so that the projections of some cutting points are as shown in Fig. \ref{AFL_Points1}. Here we have ($|M_i-K_j|$ means the difference between the coordinates of the points $M_i$ and $K_j$): 

$$
\begin{array}{ll}
|P_2-P_1|=2-i, & |L_2-L_1|=1, \\
|P_3-P_1|=3, & |L_3-L_1|=2+i, \\
|P_4-P_1|=4, & |L_4-L_1|=4, \\
|P_3-P_2|=1+i, & |L_3-L_2|=1+i, \\
|P_4-P_2|=2+i, & |L_4-L_2|=3, \\
|P_4-P_3|=1, & |L_4-L_3|=2-i.
\end{array}
$$

If we choose the cutting plane $t_1$ so that the projections of the cutting points are $L_3$ and $L_2$ positioned at the lower side of the area $A_{11}$, moving $t_1$ up along the $Z$ axis in ${\mathbb R}^3$ to a special position $t_2$ means moving $L_3$ and $L_2$ towards the upper side of $A_{11}$, so we get the points $D_3$ and $D_2$. Denoting the  appropriate projection functions with $z(t)$ and $z'(t)$ respectively, the movement of the above points result in the following functions: $z(t)-z'(t)$, $t_1<t<t_2$. We can consider the movement of these points in $A_{11}$ as a function $z(t)-z'(t)=(1-t)+i$, $t_0<t<1$ getting similar function with transformed variables (note that the points $P_2, L_3$ are lower or higher than other points, so after moving they go in or get out of the respective area $A_{i}$ as $D_3$, but we consider them as in the same area).

This process can be applied to any pair of points in each separate area $A_k$ getting the following formulae:

$$
z_j(t)-z'_j(t)=c_i+\epsilon_1\cdot t+\epsilon_2\cdot i,
$$

$$
\frac{\frac{d}{dt}{(z_j(t)-z'_j(t))}}{(z_j(t)-z'_j(t))}=\frac{\epsilon_1}{c_i+\epsilon_1\cdot t+\epsilon_2\cdot i},
$$

where $c_i\in\{1,2,3,4\}$, $\epsilon_1\in\{-1,+1\}$, $\epsilon_2\in\{-1,0,+1\}$.

These equations do not hold for the points in the areas $A_5,A_6,A_9,A_{10}$ because in these regions the arcade is alternating from one side of the faden to another.
By similar considerations, we should consider only those pairs of points $(M_i,M_j)$ where one of them lies on the arcade and the relative positions of such points can be calculated by the following formula:

\centerline{
$
z_j(t)-z'_j(t)=\epsilon_1\cdot i\cdot(\epsilon_2-t),
$
}

$$
\frac{\frac{d}{dt}{(z_j(t)-z'_j(t))}}{(z_j(t)-z'_j(t))}=\frac{-\epsilon_1\cdot i}{\epsilon_1\cdot i\cdot(\epsilon_2-t)},
$$

where $\epsilon_1\in\{-1,+1\}$, $\epsilon_2\in\{0,+1\}$.

Obviously, for the integral part in the Kontsevich formula the following equations hold:

$$
\int\limits_{0<t_1<\cdots<t_m<1}\sum_{(z_i,z'_i)\in P}(-1)^{\#P\downarrow}D_P\bigwedge\limits_{j=1}^m\frac{d(z_j-z'_j)}{z_j-z'_j}=
$$

$$
\sum_{D_l\in D^m}D_l\sum_{H_k\rightarrow D_l}(-1)^{\#P\downarrow}\int\limits_{0<t_1<\cdots<t_m<1}\bigwedge\limits_{j=1}^m\frac{d(z_j-z'_j)}{z_j-z'_j}.
$$

Hence, we get the following scheme for the computation of the Kontsevich integral:

{\bf Input:} AFL representation of a knot and an $m$-chord diagram $D_l$;\\ \\
1. Fix all the sets of points $H_k=\{(z_i,z'_i)\}$ defining $D_l$;\\
2. Calculate  as above $(-1)^{\#P\downarrow}\int\limits_{0<t_1<\cdots<t_m<1}\bigwedge\limits_{j=1}^m\frac{d(z_j-z'_j)}{z_j-z'_j}$ \\
and sum up the results.

The above integrals can be computed by standard methods.

It is clear that the computational complexity of this method depends on the number of the point sets $H_k$ defining the given chord diagram. Using standard combinatorial methods one can easily compute the number of such sets that is by far less than the number of summands in the standard formula of the Kontsevich integral. On the other hand, the functions to be integrated are very simple that makes the computation much easier.

\section{Conclusions}
In this article we have shown how an old idea can be used to develop efficient algorithms to solve actual mathematical problems. In particular, we re-introduce the AFL representation of knots introduced by Kurt Reidemeister and develop efficient algorithms to find the holonomic representation of knots introduced by Vassiliev in the late 1990s and to compute the rational factors of the Kontsevich integral for knots. It is the author's hope that the methods described above will open new perspectives in the development of fast algorithms in this field.

\end{document}